\documentclass{ws-p8-50x6-00}
\newcommand{\fcc}{$\mathrm{F}_{2}^{c\bar{c}}~$}
\begin{document}

\title{Measurement of charm production in deep inelastic scattering with the ZEUS detector}

\author{Sven Schagen \\
on behalf of the ZEUS Collaboration}

\address{NIKHEF-Amsterdam The Netherlands}

\maketitle

\abstracts{
We present measurements of charm production in DIS using the ZEUS detector. Data with $\int \mathcal{L}dt$ = 83 pb$^{-1}$ have been analysed. For the channel $D^{*+} \to D^{0}\pi_{s}^{+} \to K^{-}\pi^{+}\pi_{s}^{+} (+ c.c.)$ a cross section has been extracted, differential in the kinematical variables $Q^{2}$ and Bjorken $x$. In addition the decay $\overline{c}q \to e^-\overline{\nu}_e X$ has been studied in a data sample of $\int \mathcal{L}dt$ =  34 pb$^{-1}$. This results in a cross section, differential in $Q^2$,$x$ and $W$ of the event and in $p_{T}$ and $\eta$ of the decay electron. The structure function \fcc has also been determined for this channel. All measured cross sections show good agreement with NLO pQCD predictions from HVQDIS.
}

\section{Introduction}\label{sec:intro}
For deep inelastic electron-proton scattering perturbative QCD predicts that heavy quarks will mainly be produced by the boson-gluon fusion process: a photon emitted by the electron interacts with a gluon inside the proton to produce a $q\overline{q}$-pair, i.e. $\gamma g \to c \overline{c}$. 
The HVQDIS\cite{hvqdis} program uses NLO calculations in the DGLAP scheme at fixed order in $\alpha_s$, assuming three active flavours in the proton\cite{el}. The charm-quark is then only produced by the boson-gluon fusion.

\section{Analysis of the decay chain $D^{*+} \to D^{0}\pi_{s}^{+} \to K^{-}\pi^{+}\pi_{s}^{+} (+ c.c.)$}\label{sec:dstar}

\begin{figure}[!h]
\begin{center}
\epsfxsize=26pc
\epsfbox{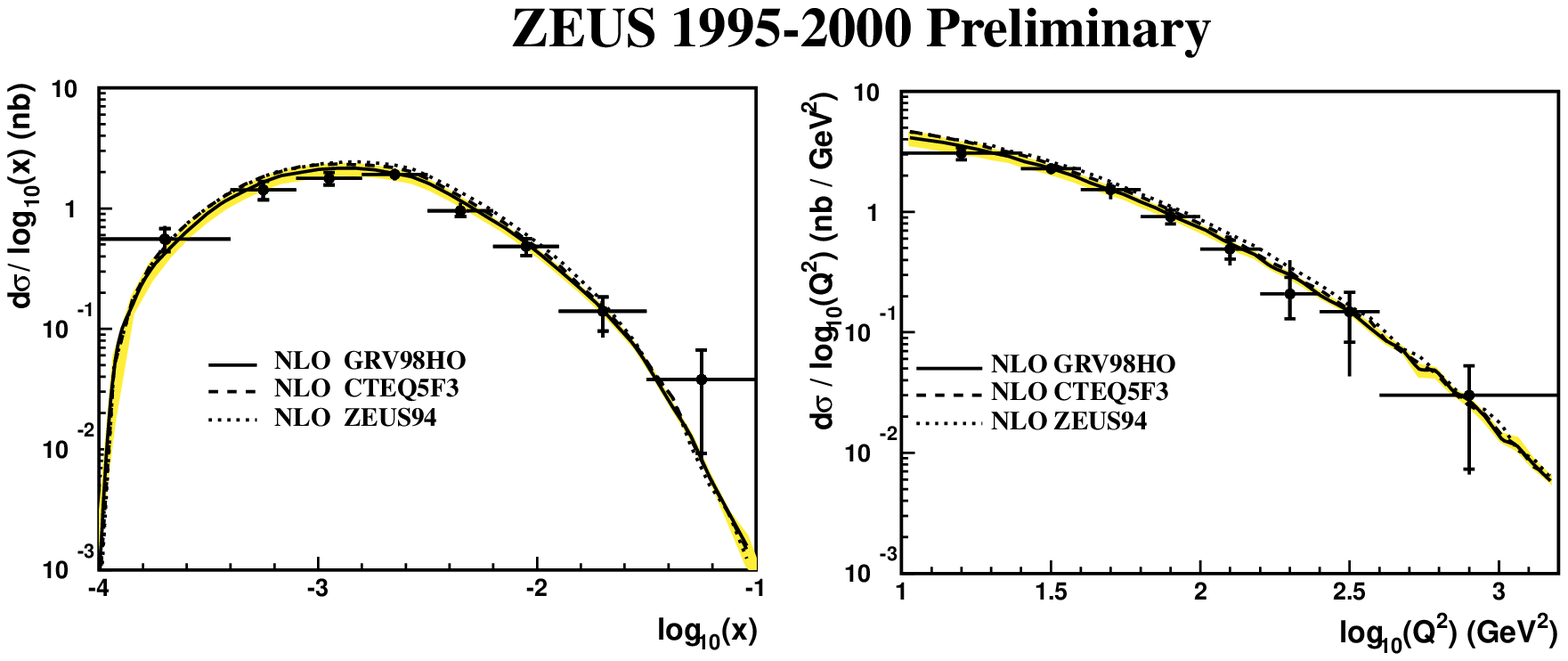}	
\end{center}
\caption{({\em left}) Charm production cross section differential in $x$ and ({\em right}) $Q^{2}$ as measured with the $D^{*}$-decay chain. The differential cross sections are compared to HVQDIS predictions for various proton structure functions sets. All of them agree with data.}\label{pic:dstar}
\end{figure}

For the analysis of the $D^*$ decay DIS events are selected that with $Q^{2}$ $>$ 10 GeV$^2$.
By combining all available fully reconstructed tracks $D^0$ and $D^{*}$ candidates are reconstructed. A clean sample of $D^*$'s can be extracted from the data by constraining the reconstructed $D^0$ mass  ($D^0$: $1.80 < \mathit{M} < 1.92$ GeV) and by cutting on the reconstructed $D^{*}$ kinematics ($D^*$: $1.5 < p_T < 15.0$ GeV and $|\eta| < 1.5$).
For the full analysis, data with an integrated luminosity of 83 pb$^{-1}$ was used. The events were collected during the 1995-1997 running period, at which HERA was operated with a 27.5 GeV positron beam and a 820 GeV proton beam, and during the 1999-2000 running period when the beam energies were 27.5 and 920 GeV, respectively. The resulting differential cross sections are shown in Fig.~\ref{pic:dstar}.

\begin{figure}[!b]
\begin{center}
\epsfxsize=24pc
\epsfbox{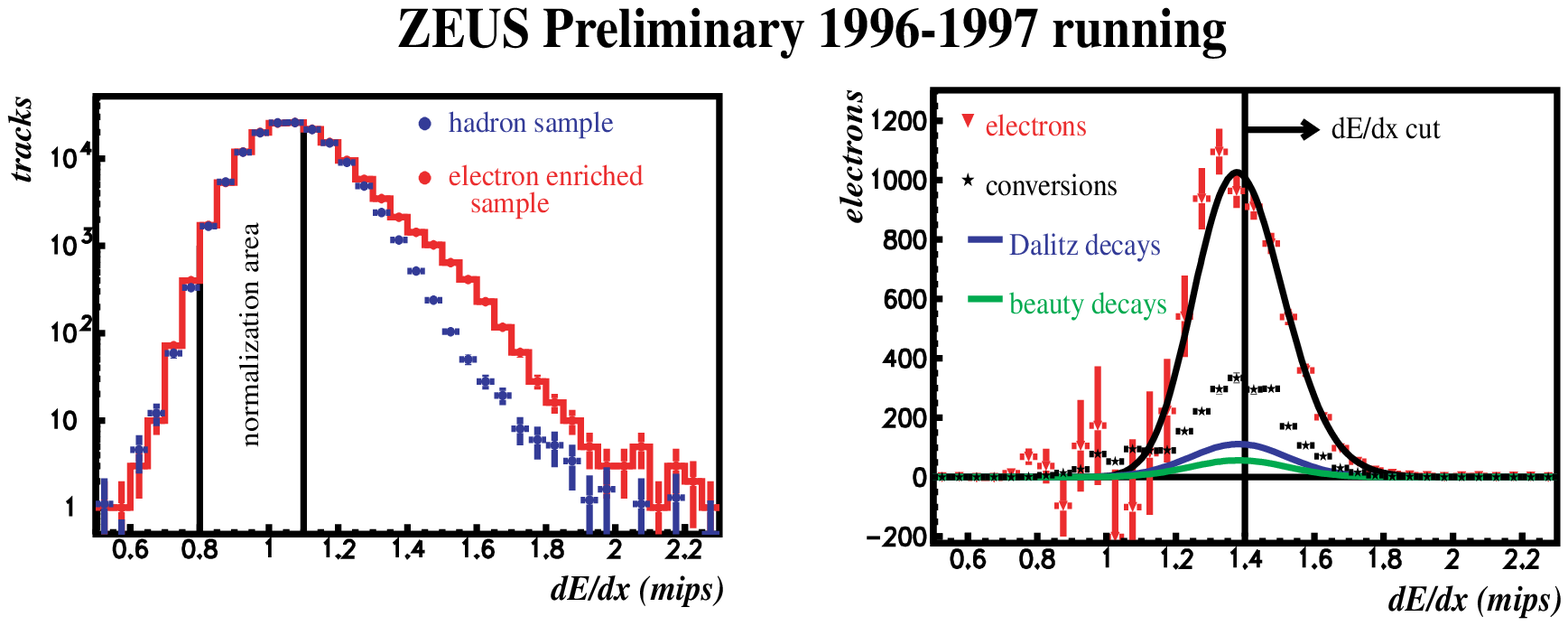}
\caption{{\em left}: The $dE/dx$ distributions for the electron enriched sample (histogram) and the hadronic sample (points).  
         {\em right}: The resulting $dE/dx$ distributions after the subtraction of the hadron distribution from the electron enriched sample with all contributions.}\label{pic:sle-signal}
\end{center}
\end{figure}

\section{Semi-leptonic decay of charmed hadrons ($\overline{c}q \to e^-\overline{\nu}_e X$)}\label{sec:sle}
\begin{figure}[!ht]
\begin{center}
\epsfxsize=23pc
\epsfbox{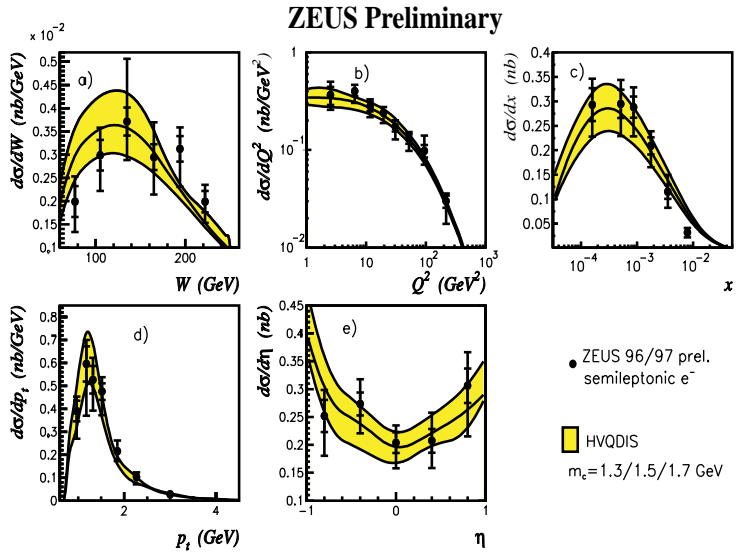}
\end{center}
\caption{Charm production cross sections differential in {\em a}) W, {\em b}) Q$^2$, {\em c}) $x$, {\em d}) $p_t^{sle}$ and {\em e}) $\eta^{sle}$ as measured through the semi-leptonic decay of charmed hadrons. The data are compared to predictions of the HVQDIS program for the GRV94HO parton distributions.}\label{pic:sle-xsecs}
\end{figure}

To study the semi-leptonic decay of charmed hadrons events with $1<Q^2<1000$ GeV$^2$ and $0.03<y<0.7$ are selected.
The electron candidates are identified based on the properties of the calorimeter cluster that is associated with them. We then consider the $dE/dx$ of all candidates (the electron-enriched sample in Fig. \ref{pic:sle-signal} ({\em left})) . The (large) hadronic background that is still within this sample is determined using the $dE/dx$ distribution of a sample containing only hadronic tracks.
After subtracting this hadronic background from the electron-enriched sample, a clean electron signal is visible (Fig. \ref{pic:sle-signal} ({\em right})). This distribution still contains electrons coming from photon conversions, Dalitz decay of the $\pi^0$ and semi-leptonic decay of beauty. These contributions are all subtracted from the sample.
For the 1996-1997 data from ZEUS (integrated luminosity of 34 pb$^{-1}$), the differential cross sections, as shown in Fig. \ref{pic:sle-xsecs} are obtained. The measurement shows good agreement with the predictions from the HVQDIS program. 
In addition, the charm structure function has been unfolded from the cross section differential in $Q^2$ and $x_{BJ}$. These results are compared to previously published ZEUS results from the $D^*$ analysis of the 1996-1997 data \cite{dstar}. Good agreement between the two analysis can be observed.

\begin{figure}[!ht]
\begin{center}
\epsfxsize=20pc
\epsfbox{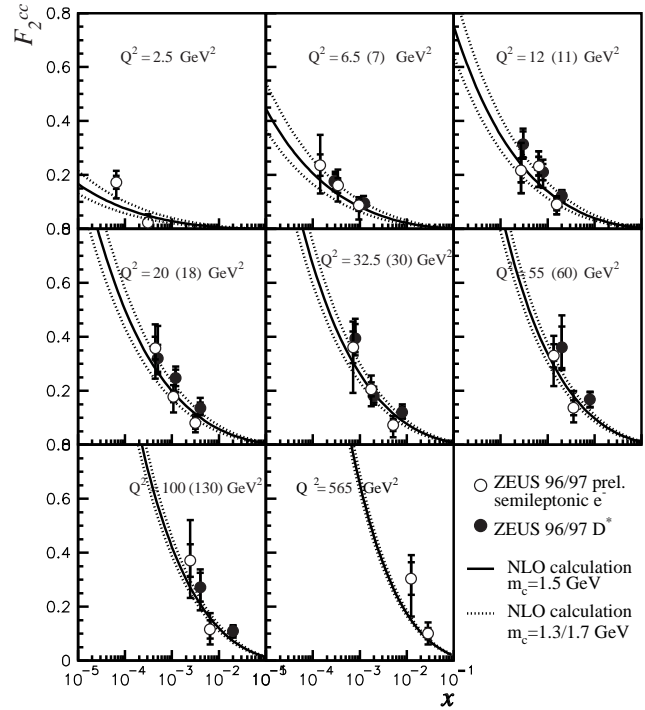}

\end{center}
\caption{The charm structure function \fcc extracted from the semi leptonic decay of charmed hadrons. The results are compared to previously published ZEUS results on the \fcc from the $D^*$ decay chain analysis. In parentheses one can find the central $Q^2$ values of the $D^*$ results.}\label{pic:sle-f2c}
\end{figure}

\section{Conclusions}\label{sec:conc}
Results on charm production in DIS using the $D^*$ meson or the semi leptonic decay into electrons have been reported. The data show good agreement with NLO DGLAP predictions for charm production through the boson-gluon fusion process as calculated by the HVQDIS program.


\begin{thebibliography}{99}

\bibitem{hvqdis}{\small B.W. Harris and J. Smith, \Journal{\PRD}{57}{2806}{1998}}

\bibitem{el}{\small E. Laenen {\em et al}, \Journal{\NPB}{392}{162}{1993}, E. Laenen {\em et al}, \Journal{\NPB}{291}{325}{1992}, S. Riemersma, J. Smith and W.L. van Neerven, \Journal{\PLB}{347}{143}{1995}}

\bibitem{dstar}{\small ZEUS Coll., {\em J.Breitweg et al}, \Journal{\em{Eur. Phys. J.} C}{12}{35}{2000}}

\bibitem{osaka_dstar}{\small ZEUS Coll., Contributed paper 451, XXXth ICHEP, Osaka, Japan, July~2000}
\bibitem{osaka_sle}{\small ZEUS Coll., Contributed paper 447, XXXth ICHEP, Osaka, Japan, July~2000} 
\end{thebibliography}
\end{document}